\documentclass{article}

\usepackage{amssymb}

\usepackage{epsfig}

\usepackage{lscape}

\usepackage{graphicx,psfrag,amsmath}

\usepackage{yfonts}

\usepackage{xcolor}

\def\t{temperature }
\def\tn{temperature}

\def\zz{neutrino }

\def\zzs{neutrinos }
\def\zzsn{neutrinos}

\def\be{\begin{equation}}
\def\ee{\end{equation}}
\def\bea{\begin{eqnarray}}
\def\eea{\end{eqnarray}}

\def\simlt{\lower.5ex\hbox{$\; \buildrel < \over \sim \;$}}
\def\simgt{\lower.5ex\hbox{$\; \buildrel > \over \sim \;$}}
\def\simpropto{\lower.2ex\hbox{$\; \buildrel \propto \over \sim
\;$}}

\newcommand{\eq}[1]{Eq\,\ref{#1}}

\newcommand{\tennd}[1]{10^{#1}}

\newcommand{\nrnd}[2]{({#1}\times{10^{#2}}\,)}

\newcommand{\tx}[1]{$t_{#1}$}
\newcommand{\txnd}[1]{t_{#1}}

\def\dtn{detection }

\def\pho{photon }

\def\phos{photons }
\def\phosn{photons}

\def\T{temperature }

\def\t{time }
\def\tn{time}

\def\yy{energy }
\def\yyn{energy}
\def\yys{energies }

\def\pls{particles }

\def\plsn{particles}

\def\prob{probability }

\def\zz{neutrino }

\def\zzs{neutrinos }
\def\zzsn{neutrinos}

\def\bhs{black holes }

\def\csss{cross-section }

\def\beq#1\eeq{\begin{equation}#1\end{equation}}
\def\beql#1#2\eeql{\begin{equation}\label{#1}#2\end{equation}}

\def\bea#1\eea{\begin{eqnarray}#1\end{eqnarray}}
\def\beal#1#2\eeal{\begin{eqnarray}\label{#1}#2\end{eqnarray}}

\def\bu{burst }

\def\bus{bursts }
\def\busn{bursts}

\def\rsh{redshift }
\def\rshn{redshift}

\def\em{emission }

\def\sp{spectrum }
\def\spn{spectrum}

\def\anphn{annihilation photon}

\def\poss{positrons }

\def\possn{positrons}

\def\anh{annihilation }

\def\abs{absorption }

\begin{document}

\title{Positron Signal from the Early Universe }

\author{Leo Stodolsky\\
Max-Planck-Institut f\"ur Physik,\\ Boltzmannstr.\,8,
85748 Garching, Germany\\
\\
Joseph Silk\\
Institut d'Astrophysique de Paris, UMR7095:CNRS \\
 UPMC-Sorbonne University, F-75014, Paris, France \\\\
 Dept. of Physics and Astronomy,
The Johns Hopkins University\\
3400 N. Charles Street
Baltimore, Maryland 21218, USA \\\\
 BIPAC, University of Oxford, 1 Keble Road,
Oxford, OX1 3RH, UK\\
} 

\maketitle

\begin{abstract}
Bursts  from the very early universe may
 lead to a detectable  signal via the production of \possn, whose
 \anh gives an observable { X}-ray signal. Using the
 absorption parameters for the 
\anphn s of 511 keV, it is found that observable  \phos would orginate at a \rsh
around  $z\approx 200--300$, resulting in   soft { X}-rays
of \yy  $\sim 2-- 3\, keV$ at present.
{  Positrons  are expected to be absent  at these  \tn s or \rshn s
  in  the standard picture of 
the early universe. Detection of the X-rays
would thus provide  dramatic support for the hypothesis of the  \busn, explosive
events at very early \tn s.} We urge the search for such  a signal. 
\end{abstract}

 We have  contemplated \cite{jl}  the possibility
 that in the very early
universe, rare but  explosive events  take place, analogous 
 to the supernovas
seen in the presently observable universe. These might be induced by such
processes as the collapse of massive  regions to \bhs
or  the formation of ``baby" \cite{Hawking:1988wm}
or ``pocket'' \cite{Guth:2000ka}
   universes'', or perhaps ``little bangs''
connected with phase transitions \cite{kort} . Although such events might  lead
to regions of spacetime which are   physically disconnected from us,
 it is plausible that during their formation, peripheral or transient
phenomena occur, as is familiar  for  the optical or  \zz bursts
 accompanying core-collapse  supernovae. Just as for the supernovae,  a quiescent remnant
may be left behind, while  a dramatic explosive effect reaches the ``outer
world''.

 The most plausible carriers of \yy or information among the presently
established
\pls would be \zzsn. Since they are neutral and have purely weak
interactions, they are the most likely to ``escape'' from the dense
environments 
of the very early universe. This would also be analogous to the
core-collapse supernovas,  where essentially all the \yy is carried away by 
\zzsn.

 Observation or \dtn  of such events would evidently  open a new chapter
in observational cosmology. 
But  there is
a great difficulty in directly detecting  such 
 early time events, namely the  high redshift  to be
anticipated. The \plsn, produced at high \rshn, will arrive to us
 with very low  \yyn,  $a (t_{em})E_{em}$, where $E_{em}$ is the \yy 
in the rest frame of their
emission, and $a(t_{em})$ is the cosmological
expansion parameter at the \em  time $t_{em}$   of the burst.
Thus \zzs emitted from an  event at cosmic time $t_{em}= 1$ second
 will have their \yys reduced by a factor $\nrnd{2}{-10}$ when arriving
at the present. Since
the \zz \csss for \dtn  is strongly \yy dependent,
 this would seem to make direct \dtn of the \bus practically impossible,  
 even if we find  \cite{jl}  that the flux  factor stops
 decreasing for very early  \em  times.

However another approach suggests itself, based on the thought that there could
be processes induced by the \bu \pls  at high \rshn, where they
are more energetic. This might    then 
lead to observable signals  \cite{jla}. The question is somewhat subtle,
since a signal created at high \rsh may not be able  to ``escape'' the dense
environments to reach the present epoch. 
 In this note we would like to briefly report on
 one of the most intriguing  and novel cases where this may be however
  possibile: the
production of \possn.

If  \zzs  with MeV \yys or more 
  arrive to the  recombination epoch or later, 
the production of positrons, whose  subsequent
  annihilation  gives 
  an  observable soft { X}-ray signal, is possible.

That is, we consider 
\beql{nuint}
\overline{\nu} +p \to e^+ + n~~~~~~~~~~~~~e^++e^-\to \gamma + \gamma 
\eeql
 giving  511 keV \phosn, whose \rsh to the present  gives  
a  soft {  X}-ray, as we shall explain.

The \dtn of such a signal would be very 
 characteristic for \busn. {  We shall
 see that the relevant \rsh factor $z$
for the origin of  observable 
{ X}-rays is on the order of some hundreds. However the \T of the universe
at such times is well below an eV, as may be judged from the fact that already
at the much earlier \t around recombination  with  $z \sim \tennd{3}$ ,
 the \T has already fallen to the order of the binding \yy of the
hydrogen atom, or some eV's.  Thus with
only sub-eV   \yys available  
 in the purely thermal equilibrium
picture, positrons should be entirely absent.}

In principle, there are other, presumably subdominant, reactions which can lead to
positrons via decay chains, such as $\overline{\nu_\mu}\, p\to \mu^+ n$,
 { followed
by $ \mu^+\to e^+ \nu \overline{\nu}$. There can also be the
production of $\pi$ mesons via hadron channels,
 as in $\nu \, p\to e^- \Delta^{++} $ followed by $\Delta^{++} \to \pi^+ p$,
where $\Delta$ is the resonance with mass 1240 MeV and the pion then decays to
a muon.  An amusing  possibility in this connection would be the
production of $\pi^o$ mesons as in $\Delta^+\to  \pi^o p $. The pion will
give two ``prompt'' { 70 MeV} \phosn, which would profit from the higher
transparency for high energy \phosn.} However, all such channels
 should be
subdominant due to the fact that, in addition to { the complicated branching
fractions involved, they have much higher \yy thresholds than the $\sim 1$
MeV for \eq{nuint}. For the muon reaction this will be near the muon mass
or  110 MeV and for
the $\Delta$ reactions around 400 MeV.} Due to the great \rsh
for the \busn, we expect  the incoming \zz \sp to be strongly peaked
towards low \yyn, making higher threshold processes subdominant \cite{pi0}.

  In principle there is also a 2.2 MeV $\gamma$
signal from capture
of the neutron on hydrogen, but due to the low matter
 density at early \tn s the 14
minute decay of the neutron is much faster.

The quantity of interest is $N_\gamma(\txnd{now})$, the present density
of \anphn s, which  gives our \dtn rate. In particular we are interested
in  its \yy  \sp
$\frac{dN_\gamma(\txnd{now})}{d\omega}$,
where $\omega$ is the present \yy of an \anphn.

Although the annihilation of a positron on a stationary electron
leads to a ``line"  at 511 keV, various effects lead to a spread
of the \pho \spn. There is the thermal motion of the atoms and the
motion of the  bound electron in the atom. However, the most important effect 
in our present problem will be due to the spread of the \rshn s
in the origin of the gamma ray. There is an essentially one-to-one
relation between the \rsh of origin $z$ and $\omega$
\beql{oto}
\omega = (1/z) \,  511 keV ~~~~~~~~~~~~~~~~~~~~~~
~~~~~d\omega = (2/3) z^{1/2} \frac{dt}{\txnd{now}} \,  511 keV.
\eeql
Since we will find that the relevant times are well into 
   the matter dominated epoch, for z we use  
$z=(\txnd{now}/t)^{2/3}$, with $\txnd{now}=\nrnd{2.9}{17} seconds$ \cite{jl}. 
The second relation shows how a spread in production \tn s $dt$ gives
a band of present \yys $d \omega$.

A \t interval $dt$ at \rsh $z$  thus gives a contribution to the present
 density of the \anphn s with \yy $\omega = (1/z) \,  511 keV$
\beql{pf}
dN_\gamma(\txnd{now})= C(z) dt= C(z) \frac{1}{z^{1/2}}\,(3/2)  \frac{ \txnd{now}}{511 keV} \,
 \,  {d\omega} \,
\eeql
where $C$ is the   factor for the conversion of \zzs into \phosn,
and the  propagation of the \phos  to the present \tn.
{ This will involve presently unknown features of the fluxes
from  the \busn, 
 their \sp and intensity. The most significant feature, due to the high \rsh
anticipated for the sources, should be the peaking of the \zz \sp towards
low \yyn. For our present purposes we leave questions pertaining to the
absolute magnitude of $C$ open and only consider those affecting the shape 
 of the \spn.}

The most important of these will be the factor  in $C$ for
  the absorption of the \phos during their propagation.
While of course positrons will be produced all along the path of a
\zz pulse, only those conversions  which are close enough so
 that the  annihilation \pho can ``escape'' to the present time
are potentially observable. 

The absorption is governed by an attenuation factor  $\cal A.$
  That is,
  the \prob of an emitted  \pho to reach us is
\beql{cala}
{\cal A} =exp(-\tau/\tau_o)
\eeql
 where $\tau$ is the  column density or 
``thickness''   of
 the matter traversed,   and 
$\tau_o $ is a parameter characterizing
 the matter.
 This parameter is usually given in  grams/cm$^2$ and   for 500 keV \phos
in hydrogen one has \cite{nist}
\beql{th}
\tau_o\approx 6 \,grams/cm^2 \, .
\eeql

This parameter is \yy dependent and so $\cal A.$ will be 
somewhat affected by the \rsh
of the traveling \phosn. However, most of the \abs will take place close to
the \t of production  and for the present rough estimates we will simply use
\eq{th}.

To evaluate \eq{cala} in the cosmological situation, we take the intervening
matter to be essentially hydrogen (the correction for helium is at the 10\%
level) and we  estimate the column density from the
present back to an earlier  \t $t$ or equivalently to an $a(t)$ or $z=1/a$. 
One first needs the density, which we take to be 
\beql{roo}
\rho(t)= \rho_o \bigl(\frac{t_{now}}{t}\bigr)^2\,,
\eeql
where $\rho_o$  is the present density of hydrogen and  the density
 scales as $1/a^3$, using  $a=(t/t_{now})^{2/3}$.

One thus has for the column density
\beql{roa}
\tau=\int_{t_{now}}^t \rho(t)dt
=\rho_o t_{now}\bigl(\frac{t_{now}}{t}\bigr)=\rho_o
t_{now}\times a^{-3/2}
=\rho_o\,t_{now}\times z^{3/2}\,,
\eeql
and thus for the absorption factor expressed  in terms of $z$

\beql{cala1}
{\cal A} =exp(-\frac{\rho_o\,t_{now}}{6\, grams/cm^2 }\, z^{3/2})
=exp(- (z/z_o)^{3/2})
\eeql
where we introduce the quantity
 $z_o=(\frac{\rho_o\,t_{now}}{6\, grams/cm^2)})^{-2/3}\approx 130$, which
characterises the absorption distance in terms of $z$.
We have taken $\rho_o$ at 5\% of the critical density, namely
  $\rho_o=\nrnd{5}{-31} grams/cm^3$.

 In addition   to $\cal A$ which
favors   nearby production of the \possn, there are
  countervailing
factors favoring higher \rshn, namely
the higher density of target protons for \eq{nuint} at early \tn s, and 
the  smaller downshift of the original  \zz \em  \yyn.

Two other possible factors of this type  cancel each other in their
$z$ or $a$ dependence. If $a$ is the expansion parameter at the time the
\poss are produced, then there is a factor $\sim 1/a^3$ for the dilution of
the original \zz \bu and at the same time a factor $\sim a^3$ for the
closeness { of the production \t } of  the  \anphn s to the present. This cancellation is due to the
fact that \pho and \zz densities \rsh in the same way.

The increasing density of proton targets with \rsh
    gives a factor $\sim (1/a)^3=  z^3$.
 As for the \zz \yy factor in the \csss \cite{nudat},
 this will depend  on whether we are in
the fully relativistic regime   $E^\nu> 1\, GeV$
  for the \zz \yy at production 
  where the \csss is linear with \yyn,
or at lower \yyn, where the behavior is closer to quadratic. In addition
to the absorption factor we thus
anticipate  a  power factor in the z dependence  of  $C$ with a power
 in the vicinity $4--5$ { (dashes are meant to indicate a range) .}

The combination of these  factors  with \eq{pf}  gives
 the \sp 
for the possibly  observable density of  \anphn s
$ N_\gamma(\txnd{now})$
\beql{propz}
\frac{dN_\gamma(\txnd{now})}{d\omega}\sim  z^p {\cal A} =z^p \times exp(- (z/z_o)^{3/2}) \, .
\eeql
with $p=3.5--4.5.$
  
The product of increasing and decreasing functions  leads to 
 a  peak at
some $z$, namely at $z=(\frac{2}{3} p)^{2/3}\times z_o$.
 We thus have a peak at $z_{peak}$
\beql{zpk}
z_{peak}\approx (1.8--2.1)\times  z_o \approx (230-- 280)\,,
\eeql
implying that the original 511 keV \pho appears at present  as a soft
{ X}-ray in the vicinity of 2 or 3 keV 

The distribution  is rather broad.
Rexpressing \eq{propz} in terms of $\omega$
\beql{propza}
\frac{dN_\gamma(\txnd{now})}{d\omega}\sim
 (1/\omega)^p exp(-(3.9/\omega)^{3/2})
\eeql
with $\omega$ in keV and $p$
in the vicinity 3.5---4.5 .
 Because of
 the range in \rsh { for production},
the original annihilation ``line'' has   becomes a broad 
and somewhat asymmetric ``bump''. A plot of \eq{propza} is shown in the
figure.

\begin{figure}
\includegraphics[width=\linewidth]{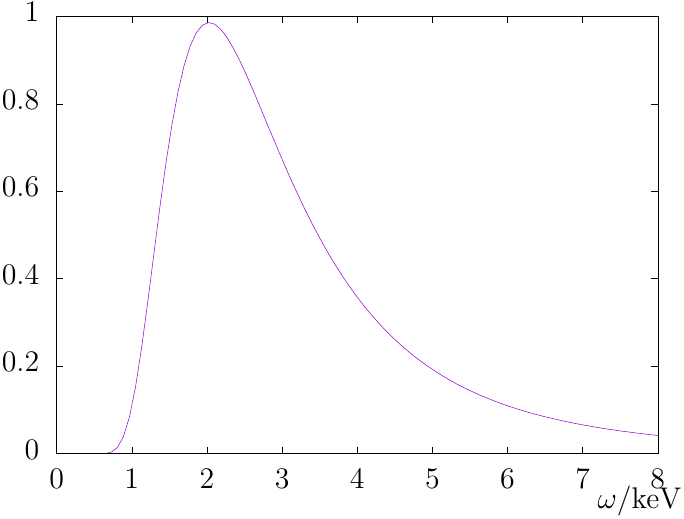}
\caption{Plot of the \pho \sp \eq{propza}. The power $p$ has been
set to $4$. The horizontal axis is in keV. Arbitrary normalization. }
\label{spec}
\end{figure}

Since a certain threshold \zz \yy is required for
 the positron production, and since a  high \rsh for the  origin of the \bus 
requires an even higher \em \yyn, there will be a limitation on how early
the \bus  can  originate. Those \bus that are emitted too early will
either have their \yys \rshn ed below threshold when arriving
at  the  production epoch,
or be absorped due to their high \yyn. Estimates \cite{jla}    using standard
\zz and early universe paramaters suggest that \zzs can ``escape''
to later \tn s when
their \em \t \tx{em}   fulfills the condition
 $\txnd{em}/sec> \nrnd{7}{-1}\,E_{em}^\nu/GeV$   (\em in the radiation dominated epoch) .
 Reexpressing this relation in terms of the \yy $E_{200}$  after the
\rsh to $z\sim 200$, one has $\txnd{em}/sec> \nrnd{6}{4}\,(E_{200}^\nu/GeV)^{2/3}$.
Thus to have the threshold \yy of $\sim 1 MeV$  at  $z\sim 200$
one finds that 
a \zz must have been emitted at earliest $\sim \nrnd{6}{2}$ seconds. Or if one asks for
an \yy of 1 GeV to give a higher production rate, the
 earliest time is about $\sim \nrnd{6}{4}$ seconds. 
 Hence  evidence for the 
  \poss would imply significant \bu
activity around these  epochs.

A question that might arise  if there is 
a flux of 511 keV \phos following  recombination is that this
 could lead to significant reionizations of the newly formed hydrogen
atoms. However we expect the \bus to be rare and 
 an estimate \cite{jla}  shows
 that their intrinsic \prob would have to be  very large
for this reionization  to be important.	
  However, should it be that
 our simple estimates are too small, then there 
could be  small ionized patches 
 after recombination.
 This could be a target for
 future generation ground-based  
  21cm and  CMB observations 
 which should be  capable of detecting patchy reionization
 \cite{Keller:2023ieg},  \cite{Bevins:2022ajf}, 
 \cite{2024A&A...681A..62M}, \cite{2017ApJ...847...64M} .

In the
 energy range under discussion, the soft { X}-ray sky is dominated by local emission.
Observed fluxes are from diffuse gas
 in  the local hot bubble, the Milky Way
 diffuse hot gas {\cite{Ueda:2022gcv}} and
  even the circumgalactic hot gas \cite{Ponti:2022nix}. { Our predicted
{ X}-ray
is an ``all-sky''effect, which could be helpful in suppressing such
backgrounds.}
  It is possible that the characteristic form of the ``bump"
in the \sp   could allow it, in refined observations,  to be picked
out among these broad backgrounds.  { Observation of this signal
would indicate the existence of explosive events in the very early universe,
not detectable in classical astronomy.} We urge looking for it.


\begin{thebibliography}{00}
\bibitem{pi0}{ If one would like nevertheless to pursue the
idea of the  $\pi^o$
channel, the methods of this note  lead to a \pho distribution
at present  with
a peak in the vicinity of 50  keV.}
\bibitem{nist} See  the ``X-ray mass attenuation tables" of NIST,
available at { physics.nist.gov/cgi-bin/Xcom/xcom3\_1 }
We consider only hydrogen, the contribution from
 helium may be seen to be small.
\end{thebibliography}
\end{document}